\documentclass[10pt,conference]{IEEEtran}
\IEEEoverridecommandlockouts

\usepackage{cite}
\usepackage{amsmath,amssymb,amsfonts}
\usepackage{algorithm}
\usepackage{algorithmic}
\usepackage{array}
\usepackage{graphicx}
\usepackage{textcomp}
\usepackage{xcolor}
\usepackage{comment}
\usepackage[font={small}]{caption}

\usepackage{url}
\begin{document}

\title{Large Language Model Partitioning \\ for Low-Latency Inference at the Edge}

\author{
\IEEEauthorblockN{
Dimitrios Kafetzis\textsuperscript{1},
Ramin Khalili\textsuperscript{2} and
Iordanis Koutsopoulos\textsuperscript{1}}
\IEEEauthorblockA{\textsuperscript{1}Athens University of Economics and Business, Department of Informatics, Athens, Greece\\
\textsuperscript{2}Huawei European Research Center, Munich, Germany}
}

\maketitle

\begin{abstract}
Large Language Models (LLMs) based on autoregressive, decoder-only Transformers generate text one token at a time, where a token represents a discrete unit of text. As each newly produced token is appended to the partial output sequence, the length grows and so does the memory and compute load, due to the expanding key-value (K/V) caches—which store intermediate representations of all previously generated tokens—in the multi-head attention (MHA) layer. As this iterative process steadily increases memory and compute demands, layer-based partitioning in resource-constrained edge environments often results in memory overload or high inference latency. To address this, aiming to reduce inference latency, we propose a resource-aware Transformer architecture partitioning algorithm, where the partitioning decision is updated at regular intervals during token generation. The approach is a myopic algorithm in the sense that it is based on instantaneously available information about device resources availability and network link bandwidths. When the algorithm is first executed, it generates a placement of blocks on devices, and in each consecutive time it is executed, it migrates these blocks among devices so that the sum of migration delay and inference delay remains low. Our approach partitions the decoder at the attention head-level, co-locating each attention head with its K/V cache and allowing dynamic migrations whenever resources become tight. By allocating different attention heads to different devices, we exploit parallel execution of attention heads and thus allow for substantial reductions in inference delays. Our experiments show that in small-scale settings (3--5 devices), the proposed method achieves within 15--20\% of an exact optimal solver’s latency, while in larger-scale tests it achieves notable improvements in inference speed and memory usage compared to state-of-the-art layer-based partitioning approaches.
\end{abstract}


\section{Introduction}
\label{section:intro}

Large Language Models (LLMs) have revolutionized a variety of natural language processing tasks, from question-answering and conversational ones to code generation. To achieve low-latency services in these applications, there is growing interest in running LLM inference at the edge, where devices such as smartphones, IoT nodes, or on-premise servers often have limited memory and compute capacity. One prerequisite for overcoming these limitations is to partition large models across multiple edge devices, harnessing collective resources instead of relying on a single node.

\begin{figure}[t]
  \centering
  \includegraphics[width=0.5\textwidth]{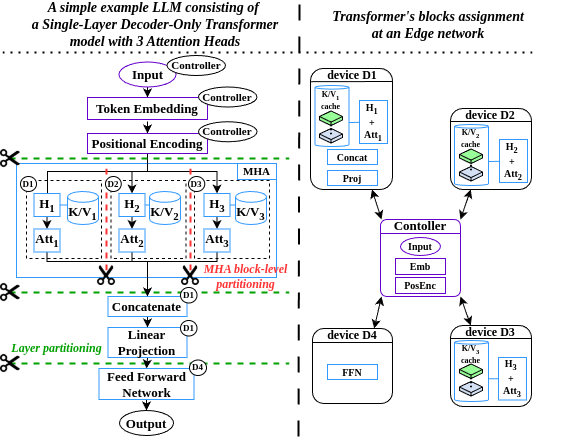}
  \caption{Illustration of our approach, showing how attention head-level partitions (red lines) and additional cuts for feed-forward network (FFN) and linear projection (proj) blocks (green lines) enable flexible assignment of blocks across multiple edge devices. The controller node makes inference requests and handles partitioning decisions, while the attention heads (with their K/V caches), FFN, and proj blocks are allocated among devices D1, D2, D3, and D4.}
  \label{fig:main_concept_toy_example}
\end{figure}

However, partitioning modern LLMs—often containing billions of parameters—remains highly non-trivial. One approach is to split the model along large block boundaries, e.g., entire decoder layers. Each decoder layer typically consists of multiple parallel attention head blocks and feed-forward networks, which can be assigned to different devices. This is what we call a coarse-grained partition, since it treats large decoder layers as a single unit. Moreover, such partitions are static in the sense that the assignment of blocks to devices remains fixed throughout the inference process \cite{zhang2024edgeshard, mudvari2024splitllm, ye2024galaxy}. While coarse-grained and static partitions can be effective for deep neural networks, large Transformer-based LLMs require more fine-grained and adaptive partitioning. This is due to their autoregressive decoding nature, where tokens are generated one by one. Specifically, as each new token is generated, the model stores intermediate representations—keys and values (K/V)—for all previously produced tokens in dedicated caches. These K/V caches appear in the multi-head attention (MHA) module of the Transformer and grow with every additional token in the sequence, since each new token must attend to all past tokens. As the inference task progresses and new tokens are generated, devices may become overloaded if each entire Transformer layer is treated as a single block.

In this paper, we focus on a \emph{single-layer decoder-only Transformer architecture} (e.g., GPT~\cite{brown2020language}), meaning it includes just one instance of the core decoder sub-modules (MHA and feed-forward blocks) rather than stacking multiple such blocks. We propose a \emph{fine-grained, attention head-level partitioning} approach that allows for more flexible use of edge resources. 

The paper contributes to state of the art as follows:
\begin{itemize}
    \item We study the problem of reducing inference delay of Transformer architectures by allocating different Transformer blocks to different devices (Figure~\ref{fig:main_concept_toy_example}).
    \item We develop an algorithm executed once before inference begins, and then at regular intervals during the generation process. When executed prior to token generation, the algorithm allocates different blocks of the Transformer architecture to different devices in order to reduce inference delay. When executed at regular intervals during token generation, the algorithm performs block migration among devices, taking migration delays into account.
    \item We demonstrate through numerical studies that our method achieves near-optimal performance in terms of inference latency reduction (within 15--20\% of an exact solver) for small networks, and achieves up to 9--10 times speedup in inference latency compared to existing partitioning approaches~\cite{zhang2024edgeshard,ye2024galaxy} in scenarios involving large numbers of generated tokens. These findings highlight the importance of attention head-level partitioning and resource management for low-latency autoregressive LLM inference at the edge.
\end{itemize}

The rest of this paper is organized as follows. Section~\ref{section:related} surveys related work on distributed Transformer inference and partitioning. Section~\ref{section:model} details our system model. Section~\ref{section:algorithm} presents our approach, and Section~\ref{section:evaluation} reports simulation results. Finally, Section~\ref{section:conclusion} concludes the paper.

\section{Related Work}
\label{section:related}

The partitioning of deep learning models across resource-constrained networks has been studied extensively to address either the training or the inference stages \cite{one6g}. A broad set of works aims to improve resource utilization, with emphasis on partitioning and parallelism techniques.

\textbf{DNN partitioning and scheduling.}
Many approaches focus on splitting DNNs across devices to leverage the heterogeneous compute and memory resources of different nodes and thereby reduce computational or communication overhead. For instance, the work in~\cite{kafetzisDNNpartitioning2024} studies DNN partitioning for inference in resource-constrained networks, considering the joint problem of partitioning many DNNs (each split between an end-device and the Base Station) and scheduling those partitions at the Base Station to minimize inference delay. Also, SplitPlace~\cite{SplitPlace2024} proposes AI-driven strategies for partitioning DNNs among edge devices based on resource availability and application demands. Similarly, the authors of~\cite{zhang2023} address the joint optimization of DNN partitioning and scheduling in end-edge-cloud systems to balance computational loads and reduce inference latency. Related lines of work couple partitioning with scheduling algorithms to improve resource-usage efficiency~\cite{yuan2023}, using deep reinforcement learning for continuous task scheduling in edge networks (i.e., tasks arrive over time and must be allocated on the fly). Overall, these methods underscore the need for adaptive (dynamic or real-time) and task-specific (i.e., tailored to each DNN-inference job) partitioning to handle limited memory and compute resources effectively.

\textbf{Pipeline and parallelism.}
A separate body of work focuses on pipeline- and parallelism-based solutions. Techniques such as GPipe~\cite{huang2019gpipe} and PipeDream~\cite{narayanan2019pipedream} accelerate model training via pipeline parallelism, splitting large models across accelerators and mitigating pipeline stalls. Extensions like AutoPipe~\cite{autopipe2021} and PipeDream-2BW~\cite{narayanan2021memory} further optimize memory footprints by applying pipeline-parallel training. On the other hand at the inference stage, Galaxy~\cite{ye2024galaxy} adopts tensor and sequence parallelism for Transformer inference at the edge. Meanwhile, long-sequence parallelism techniques~\cite{zheng2023parallelism} are designed for large input sequences and manage them using data sharding (dividing input tokens among multiple devices) or pipeline sharding (distributing computation stages).

\textbf{Inference for Transformers.}
Beyond training, methods for distributed Transformer inference have also gained traction. SplitLLM~\cite{mudvari2024splitllm} studies collaborative cloud-edge inference through device selection and model partitioning, yet it focuses primarily on single-shot inference (i.e., processing an input batch in one pass) without explicitly tackling the token-by-token memory demands of autoregressive generation. EdgeShard~\cite{zhang2024edgeshard} distributes large Transformers into shards (subdivisions of the model) but relies on a layer-wise design (i.e., treating each entire decoder layer as a single block). This approach may cause stalls and does not address the growing K/V caches, which store the key-value vectors for each newly generated token, enabling subsequent tokens to attend to them without recomputing at every decoding step.

In contrast, our approach partitions the Transformer architecture at the level of an attention head block, co-locates caches with attention head blocks to reduce communication load, and accommodates expanding memory footprints as tokens are generated. In Section~\ref{section:evaluation}, we compare our approach to static pipeline-based partitioning, showing that our proposed attention head-level partitioning outperforms those methods for growing K/V caches.

\section{System Model and Problem Formulation}
\label{section:model}

In this section, we formalize how an inference request for a single-layer, decoder-only Transformer is processed across a network of heterogeneous devices in an edge environment.

\subsection{Inference Request and Autoregressive Generation}
We consider a single inference request for text generation that is placed as an input to a device, initiated by a text of length $L_0$ tokens. During the inference, the model generates a sequence of up to $N$ tokens (units of text) one by one. At step $n \leq N$, the previously generated tokens $\{1,\ldots,n-1\}$ are fed into the decoder-only Transformer, along with the input tokens, to produce the $n$-th token. We denote by $L_n = L_0 + n$ the total sequence length in tokens at step $n$.

\paragraph{Toy Example}
Consider a simple text generation task in which the initial text is ``\texttt{The cat}'' (i.e., $L_0=2$). At step $n=1$, the model might output ``\texttt{sat},'' extending the sequence to ``\texttt{The cat sat},'' so $L_1=3$. At $n=2$, it might produce ``\texttt{on},'' yielding ``\texttt{The cat sat on}'' ($L_2=4$).  

\subsection{Devices, Memory, and Compute Capacities}
\label{subsec:devices}
Let $G = (\mathcal{V}, \mathcal{E})$ denote the edge network, where $\mathcal{V}$ is the set of heterogeneous devices and $\mathcal{E}$ is the set of communication links between them. Among these devices, one node acts as a controller, gathering resource state information from the rest. 
The inference steps are divided into intervals of size $\lambda\geq 1$, where $\lambda$ is an integer and represents the number of tokens generated per interval. These intervals are indexed by integers $\tau\in\{1,\ldots,T\}$ with $T$ representing the total number of intervals during the inference. At each interval $\tau$, the controller collects the following information from every $j \in \mathcal{V}$:
\begin{itemize}
    \item Available memory $M_j(\tau)$ (bytes),
    \item Max compute capacity $W_j$ (FLOPs/sec),
    \item Available compute capacity $C_j(\tau) \le W_j$ (FLOPs/sec), due to concurrent background processes,
    \item Link bandwidths $R_{j,k}(\tau)$ (bytes/sec) for communication with every $k \in \mathcal{V}$.
\end{itemize}

In this work, we assume that the memory and compute resources available at a device, as well as the link bandwidths, remain constant during an interval, or that these values represent the average predictions for the interval.
The controller then decides how to allocate (or migrate) the model's blocks to the devices, as will be explained in Section IV.

\subsection{Single-Layer Decoder-Only Transformer Blocks}
\label{subsec:blocks}

\textit{Definition of Blocks and Resource Requirements:}
We consider a single-layer, decoder-only Transformer architecture. The layer consists of:
\begin{itemize}
  \item $\mathcal{H} = \{1,\ldots,h\}$, the set of attention head blocks in the multi-head attention (MHA) module,
  \item a feed-forward network block, $ffn$,
  \item an output projection block, $proj$.
\end{itemize}
Hence, we define $\mathcal{B} \;=\; \mathcal{H} \;\cup\;\{ffn\}\;\cup\;\{proj\}$ as the set of blocks in the layer.
For each attention head $i \in \mathcal{H}$, there is an associated K/V cache whose size grows token by token. Specifically, at an interval $\tau$, each attention head $i$ stores more keys/values than it did at the previous interval. If $i \notin \mathcal{H}$, then $i$ is either $ffn$ or $proj$, each having its own memory demands.
For each block $i \in \mathcal{B}$, we define: 
\begin{itemize}
\item $m_i(\tau)$ as the maximum memory requirement (bytes), and 
\item $b_i(\tau)$ as the maximum compute requirement (FLOPs).
\end{itemize}
Concretely, if $i \in \mathcal{H}$, $m_i(\tau)$ represents the memory footprint of the K/V cache of attention head $i$ plus its parameters at interval \(\tau\). If $i = ffn$ or $proj$, $m_i(\tau)$ is the memory needed to store that block's parameters. Because of the model’s autoregressive nature, $m_i(\tau)$ and $b_i(\tau)$ increase with $\tau$ (reflecting the fact that, as inference process progresses, more tokens are generated and added to the K/V cache of each attention head). 

\subsection{Placement (Allocation) of Blocks}
\label{subsec:placement}
We define a binary variable $x_{ij}(\tau)$ that indicates whether block $i$ is allocated on device $j$ at interval $\tau$:
\begin{itemize}
  \item If $i \in \mathcal{H}$ (i.e., an attention head),
    \[
       x_{ij}(\tau) \;=\;
         \begin{cases}
           1, \quad \parbox[t]{5.3cm}{if attention head $i$ and its K/V cache are running at device $j$ at $\tau$}, \\
           0, \quad \text{otherwise}.
         \end{cases}
    \]
  \item If $i\notin \mathcal{H}$ (i.e., $i = ffn$ or $proj$),
    \[
       x_{ij}(\tau) \;=\;
         \begin{cases}
           1, \quad \parbox[t]{5.3cm}{if block $i$ is running on device $j$ at $\tau$}, \\
           0, \quad \text{otherwise}.
         \end{cases}
    \]
\end{itemize}
In all cases, we require that one block is placed on one and only one device at each interval $\tau$:
\[
  \sum_{j\in\mathcal{V}} x_{ij}(\tau) \;=\; 1, 
  \quad 
  \forall \, i \in \mathcal{B}.
\]

\subsubsection{Memory Capacity Constraint}
At any interval $\tau$, the total memory usage of blocks assigned to device $j$ must not exceed $M_j(\tau)$. Thus,
\begin{equation}
   \sum_{i\in\mathcal{B}} m_i(\tau)\,x_{ij}(\tau)
   \;\;\le\;\; M_j(\tau),
   \quad \forall j,\; \forall \tau.
   \label{eq:mem-constraint}
\end{equation}
Multiple blocks can be co-located on a device $j$ if their combined memory fits within $M_j(\tau)$.

\subsubsection{Migration Cost}
If a block $i$ migrates from device $j$ to device $k$ at an interval $\tau$, it incurs a migration delay:
\begin{equation}
   D_{\mathrm{mig}}^i \bigl(j \to k,\,\tau\bigr)
   \;=\;
   \frac{m_i(\tau-1)}{R_{j,k}(\tau)},
   \label{eq:mig-delay}
\end{equation}
where $m_i(\tau-1)$ is the memory footprint of block $i$ (including its K/V cache, if $i$ is an attention head) at previous interval, and $R_{j,k}(\tau)$ is the available bandwidth on link $(j,k)$ at the current interval. 

\paragraph{When migration occurs}
A migration $j \to k$ for block $i$ means $x_{ij}(\tau-1)=1$ and $x_{ik}(\tau)=1$. We generally allow at most one migration per attention head per-interval to avoid back-and-forth overhead.

\paragraph{Remark -- Why we consider a single-layer decoder}
In our model, we focus on a single decoder layer to show fine-grained partitioning at the attention head-level and dynamic assignment of the partitioned blocks. This choice keeps the modeling complexity manageable for a proof-of-concept. Although we model a single-layer decoder for clarity, our attention head-level partitioning and migration approach can be applied independently to each layer in multi-layer Transformers.

\subsection{Communication and Processing Delays}
\paragraph{Communication Latency}
We adopt an abstract model of the single-layer decoder:
\begin{itemize}
    \item At interval $\tau$, each attention head $i\in\mathcal{H}$ produces output that must be sent to $proj$ block.
    \item Then $proj$ transmits its output to $ffn$ block.
\end{itemize}

If, for a particular head $i$, the head is on device $j$ and $proj$ is on device $k$, the latency to transfer the data $W_{i\to proj}(\tau)$ is:
\begin{equation}
    D_{j\to k}^{\,i}(\tau) \;=\; \frac{W_{i\to proj}(\tau)}{R_{j,k}(\tau)},
    \label{eq:head-proj}
\end{equation}
and if $proj$ resides on $k$ while $ffn$ is on $\ell$, we have
\begin{equation}
    D_{k\to \ell}^{\,proj}(\tau) \;=\; \frac{W_{proj\to ffn}(\tau)}{R_{k,\ell}(\tau)}.
    \label{eq:proj-ffn}
\end{equation}

When multiple blocks on the same device $j$ send outputs to the same device $k$, they must share the outgoing link $(j,k)$. 
We assume that these transmissions are performed in series. 

\paragraph{Processing Delay}
For block $i$ at $\tau$, if $i$ is on device $j$, we define
\begin{equation}
    D_{ij}(\tau)
    \;=\;
    \frac{b_i(\tau)}{C_j(\tau)},
    \label{eq:processing-delay}
\end{equation}
as the time to process $b_i(\tau)$ FLOPs, given available compute $C_j(\tau)\le W_j$. However, if multiple blocks (e.g., several attention heads) share device $j$ concurrently, they also share $C_j(\tau)$. One simple scheduling policy is to process them sequentially on $j$, in which case we sum their compute demands; though more sophisticated concurrency models can be used as needed.

\subsection{Total Inference Delay at $\tau$}
\label{subsec:Dt}
Suppose we fix a placement $\mathcal{A}(\tau)$, i.e.\ $x_{ij}(\tau)$ for all $i\in\mathcal{B}$ at interval $\tau$, where $j$ indexes the devices. Then $d(\tau,i)$ denotes the device hosting block $i$. The total inference delay $D_{\mathrm{T}}(\tau)$ can be decomposed as follows, adapting the standard decoding pipeline (input $\to$ attention heads $\to proj \to ffn$):
\begin{equation}
\begin{split}
  D_{\mathrm{T}}(\tau)
    &= \max_{i \in \mathcal{H}}
         \Bigl\{\,D_{\mathrm{in}\to d(\tau,i)}^{\mathrm{in}}(\tau)
               + D_{i,d(\tau,i)}(\tau) \\
    &\quad+ D_{d(\tau,i)\to d(\tau,proj)}^{\,i}(\tau)\Bigr\}
         + D_{\,d(\tau,proj)\,\to\,d(\tau,ffn)}^{\mathrm{proj}}(\tau),
\end{split}
\label{eq:total-inference-delay}
\end{equation}
where:
\begin{itemize}
    \item $D_{\mathrm{in}\to d(\tau,i)}^{\mathrm{in}}(\tau)$ is the delay of moving input tokens from their initial storage (e.g.\ a controller node) to the attention head’s device $d(\tau,i)$;
    \item $D_{i,d(\tau,i)}(\tau)$ is the processing delay of attention head $i$ on its device $d(\tau,i)$ (per \eqref{eq:processing-delay});
    \item $D_{d(\tau,i)\to d(\tau,proj)}^{i}(\tau)$ is the communication delay from attention head $i$ to $proj$ (per \eqref{eq:head-proj});
    \item $D_{\,d(\tau,proj)\,\to\,d(\tau,ffn)}^{\,proj}(\tau)$ is the delay from $proj$ to $ffn$ (per \eqref{eq:proj-ffn}).
\end{itemize}

When multiple blocks share a device or link, there are concurrency effects:
\begin{itemize}
    \item \emph{Compute concurrency:} If multiple heads or other blocks run on the same device $j$ and share compute $C_j(\tau)$, the processing times will depend on the block compute scheduling policy.
    \item \emph{Link concurrency:} If multiple heads on the device $j$ send outputs to the device $k$ simultaneously, the transfer delays in $D_{d(\tau,i)\to d(\tau,proj)}^{i}(\tau)$ may be aggregated or scheduled.
\end{itemize}
Equation~\eqref{eq:total-inference-delay} retains a simplified pipeline form for clarity.

\subsection{Decision at Each $\tau$}
\label{subsec:decision-step}

We must decide a new placement $\mathcal{A}(\tau)$, i.e. a mapping $\{i \mapsto d(\tau,i)\}$ for each block $i\in\mathcal{B}$ at each interval $\tau$. If block $i$ was on device $d(\tau-1,i)$, but we now place it on $d(\tau,i)\neq d(\tau-1,i)$, that constitutes a migration with cost $D_{\mathrm{mig}}^i\bigl(d(\tau-1,i)\,\to\,d(\tau,i),\,\tau\bigr)$ per \eqref{eq:mig-delay}.
We sum the delays of these migrations:
\begin{equation}
    D_{\mathrm{mig}}^{\mathrm{total}}(\tau)
    \;=\;
    \sum_{i\in\mathcal{B}}
    D_{\mathrm{mig}}^i \bigl(d(\tau-1,i)\,\to\,d(\tau,i),\,\tau\bigr),
    \label{eq:mig-total}
\end{equation}
assuming migrations happen sequentially or that no two or more blocks migrate over the same link simultaneously.

\paragraph{Objective}
At each $\tau$, we need to find an assignment $\mathcal{A}(\tau)$ in order to minimize $D_{\mathrm{T}}(\tau) \;+\; D_{\mathrm{mig}}^{\mathrm{total}}(\tau)$,
subject to the memory constraint~\eqref{eq:mem-constraint}, and the network’s communication/processing costs. At $\tau=1$, we initialize $\mathcal{A}(1)$ to minimize $D_{\mathrm{T}}(1)$, while for $\tau>1$ we also include the migration cost $D_{\mathrm{mig}}^{\mathrm{total}}(\tau)$.

\begin{figure}[t]
    \centering
    \includegraphics[width=0.44\textwidth]{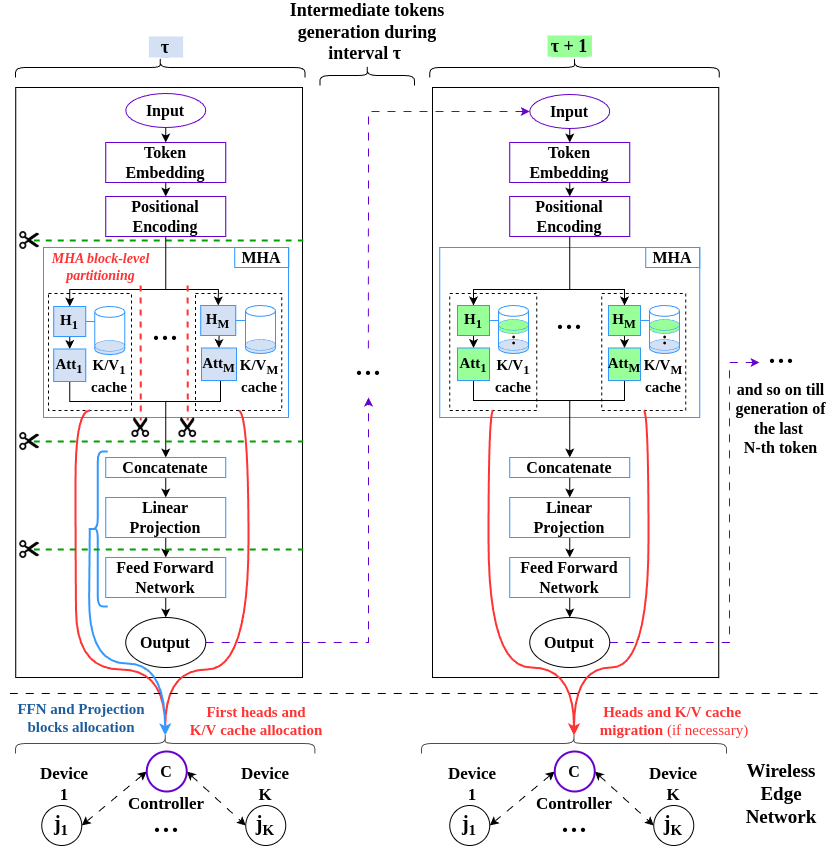}
    \caption{A single-layer decoder-only Transformer at intervals $\tau$ and $\tau + 1$, continuing autoregressively up to the $N$-th token generation. Red lines show attention head-level partitioning in MHA; green lines indicate projection and feed-forward partitioning. Cyan vs.\ green blocks highlight new tokens building on the K/V caches of the attention heads. At the bottom, a controller allocates these blocks across edge devices.}
    \label{fig:main_concept}
\end{figure}

\subsubsection{Per-Interval Assignment Policy (Myopic Algorithm)}
\label{subsec:myopic}
The controller node executes a centralized algorithm that decides the assignment policy (Figure~\ref{fig:main_concept}) i.e., the block-to-device mapping $\mathcal{A}(\tau)$ at each interval $\tau$ as follows:
\begin{itemize}
    \item The controller receives updated $\{C_j(\tau)\}$, $\{M_j(\tau)\}$, and $\{R_{j,k}(\tau)\}$ from all devices/links.
    \item It observes the previous assignment $\mathcal{A}(\tau-1)$.
    \item It computes the new assignment $\mathcal{A}(\tau)$ by solving a constrained optimization that minimizes $D_{\mathrm{T}}(\tau) + D_{\mathrm{mig}}^{\mathrm{total}}(\tau)$ under the memory constraint \eqref{eq:mem-constraint}.
    \item It performs any required block migrations.
\end{itemize}
If we had a priori knowledge (or accurate predictions) of the capacity evolutions $\{C_j(\tau), M_j(\tau), R_{j,k}(\tau)\}$ on each $\tau$, then we would solve for a global schedule $X = \{X(1),\ldots,X(T)\}$ where each $X(\tau)$ is a matrix $\mathcal{B}\times\mathcal{V}$ with $x_{ij}(\tau)=1$ indicating that block $i$ runs on device $j$ during the interval $\tau$. In this work, we focus on the simpler myopic scheme that takes as input the current block allocation and the current availability of resources in memory, compute and bandwidth and performs the migration that gives the best cost (migration plus inference) as perceived at the next interval. 
This approach is more realistic, since it is more plausible to know the instantaneous bandwidth, memory, and compute resource amounts and decide based on them on what the best next move is, rather than knowing the entire process of evolution of these resources.

In the next section, we present our proposed practical heuristic of this placement and migration procedure.

\section{Resource-Aware Algorithm for LLM Block Assignment to Devices}
\label{section:algorithm}

In this section, we present a practical heuristic for the myopic assignment procedure described above. Recall that the controller that executes the centralized heuristic algorithm at each interval $\tau$, knows the device compute capacities $\{C_j(\tau)\}$, memory capacities $\{M_j(\tau)\}$, bandwidths $\{R_{j,k}(\tau)\}$, and the previous placement $\{x_{ij}(\tau-1)\}$. At each $\tau$, the controller seeks to find a block-to-device mapping $\mathcal{A}(\tau) = \{ x_{ij}(\tau) : i\in \mathcal{B},\, j \in \mathcal{V} \}$, to minimize the interval’s total delay (inference plus migration), subject to the time-varying memory constraints~\eqref{eq:mem-constraint} and the placement definition from Section~\ref{subsec:placement}.

\subsection{Algorithm Overview}
\label{subsec:alg-overview}

At a high level, the algorithm steps are as follows:
\begin{enumerate}
    \item Collect resource availability from devices and links: The algorithm gathers the updated device compute capacities $\{C_j(\tau)\}$, memory $\{M_j(\tau)\}$, and link bandwidths $\{R_{j,k}(\tau)\}$, along with the previous assignment $\{x_{ij}(\tau-1)\}$. 

    \item Compute feasibility scores: For each block $i \in \mathcal{B}$ and device $j$, we calculate a scoring function $\mathcal{S}(i,j,\tau)$ to gauge the suitability of placing $i$ in $j$. This step is per block, so it does not yet consider concurrency with other blocks on $j$. A score $\mathcal{S}(i,j,\tau)\le 1$ simply indicates that $i$ by itself could fit on $j$ given $j$'s resources.

    \item Assign blocks and handle migrations:
    We pick the device $j^*$ yielding the lowest feasible score for each block $i$. If $j^*\neq j_{\text{old}}$, we incur a migration cost $D_{\mathrm{mig}}^i\bigl(j_{\text{old}}\to j^*, \tau\bigr)$ per \eqref{eq:mig-delay}.
    If no device is feasible for $i$ alone, we attempt to resolve overload by migrating other blocks from devices with insufficient memory or compute.

    \item Verify constraints or backtrack: After assigning all blocks, we check memory demands (sum of all blocks’ usage on device $j$ vs.\ $M_j(\tau)$) and compute demands (total load vs.\ $C_j(\tau)$). If any constraint is violated, we backtrack by removing a minimal set of blocks (the fewest blocks needed to resolve the violation) from the overburdened device and reassigning them. This ensures concurrency among blocks is handled collectively.

    \item Return the new placement to the controller: Once we finalize $\{x_{ij}(\tau)\}$ for all blocks $i$, the controller applies the assignment, triggers migrations if needed, and proceeds to the next interval $\tau + 1$.
\end{enumerate}
We effectively execute the algorithm at every interval $\tau$, where the size $\lambda$ of each interval $\tau$ is chosen so each interval is on the order of a few seconds, thus allowing enough time for migrations to occur.

\paragraph{Scoring Function}
\label{par:scoringfunction}
To decide on which device $j$ block $i$ should be placed at $\tau$, we define the scoring function
\[
   \mathcal{S}(i,j,\tau)
   \;=\;
   \max \Bigl\{
     \tfrac{m_i(\tau)}{M_j(\tau)},
     \,\tfrac{b_i(\tau)}{C_j(\tau)},
     \,\mathrm{CommFactor}(i,j,\tau)
   \Bigr\}.
\]
This scoring function measures how each block $i$ would use memory, impose compute load, and incur communication overhead on a candidate device $j$. Specifically:
\begin{itemize}
  \item Memory/Compute feasibility: The terms $\tfrac{m_i(\tau)}{M_j(\tau)}$ and $\tfrac{b_i(\tau)}{C_j(\tau)}$ measure whether block $i$ alone could fit into $j$’s resource capacities. A low ratio suggests device $j$ can handle $i$, assuming $j$’s resources are not already fully consumed by other blocks.
  \item Communication overhead: $\mathrm{CommFactor}(i,j,\tau)$ approximates data transfer times if $i$ must exchange information with blocks on different devices.
\end{itemize}
A device $j$ is said to be individually feasible for block $i$ if $\mathcal{S}(i,j,\tau)\le 1$. However, in the presence of multiple blocks competing for device $j$, we rely on the final constraints check (Step~4 in the algorithm in section~\ref{subsec:alg-overview}) to ensure that the sum of $m_i(\tau)$ across all assigned blocks does not exceed $M_j(\tau)$, and likewise that total compute can be scheduled within $C_j(\tau)$. If assigning $i$ to $j$ alongside other blocks pushes $j$’s memory or compute usage beyond feasible limits, we resolve this by invoking \texttt{ResolveResourceOverload} (Section~\ref{subsec:resolve-contention}) to attempt block migrations that free up memory or compute resources.

\paragraph{Termination Criteria}
To ensure the assignment procedure at each interval $\tau$ completes in finite time, we impose a runtime safeguard with two stopping conditions: \textit{(i)} a time limit $T_{\mathrm{max}}$ (i.e., a real-time upper bound on how long the controller can spend adjusting assignments), and \textit{(ii)} an iteration bound $U = |\mathcal{B}|\times|\mathcal{V}|$, which caps repeated reassignments (migrations/backtracking) so the algorithm does not endlessly retry assignments of blocks. If either limit is exceeded, the algorithm returns \textsc{Infeasible}. These conditions keep our myopic approach tractable for each interval.

\subsection{Explanation of the Main Steps of Algorithm~\ref{alg:distribution}}
\label{subsec:alg-explanation}

Algorithm~\ref{alg:distribution} runs at each interval $\tau$ and the pseudocode follows. 
\begin{algorithm}[!b]
\centering
\caption{Resource-Aware algorithm for LLM block assignment at interval $\tau$}
\begin{algorithmic}[1]
\REQUIRE Network $G=(\mathcal{V},\mathcal{E})$, set of blocks $\mathcal{B}$,\\ interval index $\tau$, previous assignment $\{x_{ij}(\tau-1)\}$, \\ iteration bound $U$, time limit $T_{\mathrm{max}}$
\ENSURE Assignment $\{x_{ij}(\tau)\}$ or \textsc{Infeasible} if no valid solution is found
\STATE Initialize $\text{migrationCount}\leftarrow 0$, $\text{backtrackCount}\leftarrow 0$
\STATE Start timer to enforce $T_{\mathrm{max}}$
\STATE \texttt{UpdateResourceUsage($\tau$)} // gather $M_j(\tau), C_j(\tau)$, and link bandwidths $\{R_{j,k}(\tau)\}$ for each device $j \in \mathcal{V}$.
\STATE Sort $\mathcal{B}$ into $blocksQueue$ (descending by $m_i(\tau)$ or $b_i(\tau)$)
\FOR{\textbf{each} block $i \in blocksQueue$}
    \STATE Compute score $\mathcal{S}(i,j,\tau)$ for all $j \in \mathcal{V}$
    \STATE Select device $j^* \gets \arg\min_{j}\mathcal{S}(i,j,\tau)$
    \IF{$\mathcal{S}(i,j^*,\tau)\le 1$}
        \STATE $x_{ij^*}(\tau)\leftarrow 1$ \quad // tentative assignment
        \STATE \textbf{if} $\sum_{i'} m_{i'}(\tau)\,x_{i'j^*}(\tau) > M_{j^*}(\tau)$ \textbf{or} $\sum_{i'} b_{i'}(\tau)\,x_{i'j^*}(\tau) > C_{j^*}(\tau)$
        \STATE \quad $x_{ij^*}(\tau)\leftarrow 0$; // undo assignment 
        \STATE \quad \texttt{ResolveResourceOverload}($i,\tau$)
        \STATE \quad $\text{migrationCount}\leftarrow \text{migrationCount}+1$
        \STATE \quad \textbf{if} $\text{migrationCount} > U$ 
               \textbf{then return} \textsc{Infeasible}
        \STATE \textbf{else if} $i$ moved from $j_{\text{old}} \neq j^*$
        \STATE \quad $\text{migrationCount}\leftarrow \text{migrationCount}+1$
        \STATE \quad \textbf{if} $\text{migrationCount} > U$ 
               \textbf{then return} \textsc{Infeasible}
    \ELSE
        \STATE \texttt{ResolveResourceOverload}($i,\tau$)
        \STATE $\text{migrationCount}\leftarrow \text{migrationCount}+1$
        \STATE \textbf{if} $\text{migrationCount} > U$ 
               \textbf{then return} \textsc{Infeasible}
    \ENDIF
    \STATE \textbf{if} $\text{elapsed time} > T_{\mathrm{max}}$ 
           \textbf{then return} \textsc{Infeasible}
\ENDFOR
\IF{\textbf{not} $\texttt{allConstraintsSatisfied}\bigl(\{x_{ij}(\tau)\}\bigr)$}
    \STATE \texttt{BacktrackForResourceViolations}($\tau$)
    \STATE $\text{backtrackCount}\leftarrow \text{backtrackCount}+1$
    \STATE \textbf{if} $\text{backtrackCount} > U$ 
           \textbf{then return} \textsc{Infeasible}
\ENDIF
\STATE \textbf{return} Full assignment matrix $\{x_{ij}(\tau)\}$
\end{algorithmic}
\label{alg:distribution}
\end{algorithm}

\begin{itemize}
    \item Lines 1--3: We reset the migration/backtrack counters, record the start time, and collect up-to-date memory and compute capacities for each device at $\tau$ and as well as the links' bandwidth.
    \item Line 4: We generate a sorted list of blocks (attention heads, feed-forward, projection, etc.) based on descending memory or compute demand, ensuring higher-demand blocks are considered first.
    \item Lines 5--22: For each block, we select the device with the lowest \(\mathcal{S}(i,j,\tau)\). We then tentatively place the block there and verify total resource usage on that device. If usage exceeds its limits, we revert the assignment and call \texttt{ResolveResourceOverload} to migrate other blocks and free capacity. Any block migration increments a counter, and exceeding its bound leads to \textsc{Infeasible}.
    \item Lines 23--29: If memory or compute constraints remain violated after assignment, we invoke \texttt{BacktrackForResourceViolations} to relocate blocks causing the conflict. If too many backtracking or migration steps occur, or if we reach the time limit, we return \textsc{Infeasible}.
    \item Line 30: Once all constraints are satisfied, the algorithm concludes by returning the final feasible mapping of all blocks to devices for the interval $\tau$.
\end{itemize}

\subsubsection{Resolving Resource Overload}
\label{subsec:resolve-contention}
If a block $i$ cannot be placed on any device without exceeding memory/compute constraints at $\tau$, \texttt{ResolveResourceOverload} tries to migrate other blocks. For instance, to move a block from device $j$ to $k$, we check that the sum of memory already on $k$ plus $m_i(\tau)$ remains within $M_k(\tau)$. Similar checks apply for compute. If no such migration fixes the overload, we escalate to \texttt{BacktrackForResourceViolations}.

\subsubsection{Backtrack for Resource Violations}
\label{subsec:backtracking}
When constraints remain violated after attempts to resolve overload, we use \texttt{BacktrackForResourceViolations} to reassign a minimal set of blocks that cause the violation. If it still fails, we return \textsc{Infeasible}.
This approach remains consistent with the myopic policy from Section~\ref{subsec:decision-step}, handling each token-generation interval in a bounded-time manner.

Overall, the worst-case time complexity of Algorithm~\ref{alg:distribution} at each interval $\tau$ is $\mathcal{O}(|\mathcal{B}|^2 |\mathcal{V}|)$. In the most demanding scenarios, each block may require checking all devices and triggering multiple reassignments to resolve resource overloads. This complexity remains practical for small- and medium-scale deployments, but may require approximation heuristics in large-scale settings.

By iterating these steps, we obtain a mapping $\mathcal{A}(\tau)$ that aims to minimize inference and migration delays at each interval $\tau$. In Section~\ref{section:evaluation}, we demonstrate the evaluation of the performance of Algorithm~\ref{alg:distribution}.

\section{Evaluation}
\label{section:evaluation}

We evaluate our resource-aware algorithm in two distinct settings to assess both correctness (optimality gap) and scalability:
\begin{itemize}
    \item Small-scale (3--5 devices, $N=4$ tokens): We use an exhaustive (exact) solver to find the optimal assignment. This is feasible only for small-scale setups. We measure our heuristic's deviation from the global optimum that is found with exhaustive enumeration of the solutions, and compare against simpler baselines—Greedy, Round-Robin, Static, and Dynamic—as described below.

    \item Medium-scale (25 devices, up to $N=1000$ tokens): We scale up the number of devices, examining whether our approach outperforms other methods under more realistic edge settings with higher heterogeneity. We benchmark our resource-aware approach against state-of-the-art partitioning frameworks (EdgeShard~\cite{zhang2024edgeshard}, Galaxy~\cite{ye2024galaxy}) for large-scale Transformer inference across multiple devices.
\end{itemize}
We execute our resource-aware algorithm at fixed intervals of size $\lambda = 1$, meaning that one token is generated during each interval $\tau$. In other words, the algorithm updates the block-to-device assignment $\mathcal{A}(\tau)$ once per token. This setup represents a worst-case scenario in terms of migration overhead, as it triggers the controller to re-evaluate and potentially migrate blocks at every single decoding step. As a result, it creates the highest possible frequency of migration and placement decisions, which stresses the system and highlights the efficiency of our method under tight constraints.

\subsection{Baseline Methods}
\label{subsec:baseline-methods}
We compare our approach to the following baselines, which illustrate different ways of Transformer architecture partitioning:
\begin{itemize}
    \item Greedy: Sort blocks in descending order of resource demand and place them on the first feasible device without re-checking feasibility in subsequent steps.
    \item Round-Robin: Assign blocks sequentially to devices in a cyclic order, ignoring different resource requirements at different stages of the token generation process.
    \item Static: Do one initial assignment for all blocks and never migrate them during token generation.
    \item Dynamic: As in Resource-Aware algorithm, re-checks assignments at each step but treat each layer as one block. 
    \item EdgeShard~\cite{zhang2024edgeshard}: Assigns entire Transformer layers to devices. This static, layer-based partitioning does not adapt to resource changes or K/V cache growth.
    \item Galaxy~\cite{ye2024galaxy}: Partitions the Transformer into contiguous layer shards for pipeline parallelism and splits each shard’s large matrix multiplications across multiple devices for tensor parallelism.
\end{itemize}

\subsection{Evaluation Setup}
\label{subsec:evaluation-setup}
We built a custom Python simulator\footnote[1]{The full codebase of our implementation is publicly available in \cite{Kafetzis2025transformerInferenceSimulator}} in a discrete-event fashion to model each token generation step. A central controller gathers devices' memory/compute availability and links' bandwidth at every interval $\tau$ and runs our resource-aware algorithm or a baseline algorithm. 
The simulator’s modular design supports custom device topologies, concurrency models, or partitioning policies, enabling easy adaptation to new distributed inference experiments.

\paragraph{Transformer Configurations}
\label{subsubsec:baseline-methods}
We focus on a single-layer decoder with $h$ attention heads and an embedding dimension $D$ that is the size of each token's representation. This parameter strongly influences both memory and compute usage. For a \emph{Large} LLM model setup ($h=32, D=2048$), we approximate GPT-2/LLaMA scales. 
The initial input text length is $L_0=64$.

\paragraph{Device Capabilities}
We sample each device's memory availability $M_j(\tau)$ (GB) and compute capacity $C_j(\tau)$ (GFLOPS) from log-normal distributions for heterogeneity (e.g., $M_j(\tau) \in [2,8]$ GB, $C_j(\tau)\in[5,50]$ GFLOPS) \cite{reiss2012google}. For the network, we assign each link $(j,k)$ a bandwidth randomly drawn from $[1,\,10]$\,Gbps, reflecting diverse edge conditions and assuming full connectivity between devices. Table I summarizes the memory and compute usage calculation formulas per-token generation step for the considered Transformer architecture blocks, derived from the analysis in \cite{vaswani2017attention}; where in these formulas, $b$ is the number of bytes for each model parameter (i.e., the numerical values learned during training that define how inputs are processed), and $d = D/h$.

\begin{table}[h]
\vspace*{0.05in}
\centering
\small
\renewcommand{\arraystretch}{1.2}
\begin{tabular}{|l|l|l|}
\hline
\textbf{Block} & \textbf{Memory ($m_i(\tau)$)} & \textbf{Compute ($b_i(\tau)$)} \\
\hline
\textbf{Attn. Head $i$} 
 & $3 L_{\tau} \, d \, b + 3 D \, d \, b$ 
 & $3 L_{\tau} \, D \, d + L_{\tau}^2 \, d$ \\
\textbf{K/V cache}
 & $m^{cache}_{i}(\tau) = \tau \, D \, b$
 & N/A \\
\hline
\textbf{Projection}
 & $m_{proj}(\tau) = L_{\tau} \, D \, b$
 & $b_{proj}(\tau) = L_{\tau} \, D^2$ \\
\textbf{FFN}
 & $m_{ffn}(\tau) = 4 \, L_{\tau} \, D \, b$
 & $b_{ffn}(\tau) = 8 \, L_{\tau} \, D^2$ \\
\hline
\end{tabular}
\caption{Resource usage formulas assuming one token is generated per interval $\tau$ i.e., assuming that one token is generated per interval, so that $n = \tau$}
\label{tab:res-usage}
\end{table}

\paragraph{Metrics}
We consider two metrics in our evaluation, inference latency and memory usage. \emph{Inference Latency} is total elapsed time to generate $N$ output tokens, encompassing computation, communication, and any migration delay. \emph{Memory Usage} can be measured as either the sum across all devices or the maximum usage on a single device, illustrating how efficiently our approach handles expanding K/V caches.

\subsection{Small-Scale Scenario Results}
\label{subsec:small_scenario}

Here, an \emph{exact} search is possible and practical so as to find the optimal assignment. The ratio of each method’s total latency to that of the optimal solution shows that our Resource-Aware approach remains within 15--20\% of optimal, while Greedy and other baselines may lag behind by 40--60\%. This underscores the impact of attention head-level partitioning for short decoding sequences for a small number of devices.

\subsection{Medium-Scale Scenario Results}
\label{subsec:medium_scenario}

Next, we simulate a network of 25 devices (2--8 GB memory, 5--50 GFLOPS compute) with $N$ up to 1000 tokens. We also inject background tasks to emulate fluctuating compute load. We compare Resource-Aware with the EdgeShard and Galaxy frameworks.

\begin{figure}[ht]
  \centering
  \includegraphics[width=0.42\textwidth]{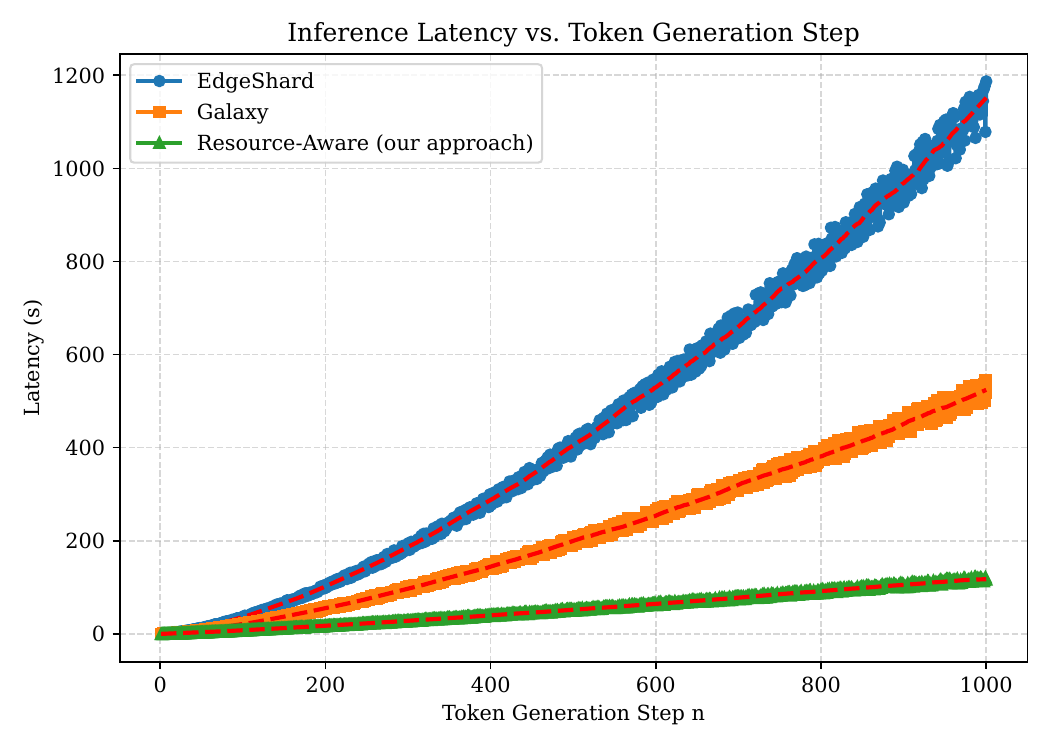}
  \caption{Inference latency vs.\ generated token step $n$ on 25 devices. Our approach (Resource-Aware) avoids steep growth.}
  \label{fig:latency_big_scenario}
\end{figure}

\paragraph{Inference Latency vs.\ Token Generation Step $n$}
Figure~\ref{fig:latency_big_scenario} shows that while all methods see rising latency as more tokens accumulate, EdgeShard eventually surpasses 1000 seconds at the last generated token, and Galaxy slows to about 400--600 seconds. In contrast, our fine-grained, attention head-level partitioning dynamically reallocates attention heads to prevent severe overload, keeping latency under 200 seconds at the last generated token.

\begin{figure}[ht]
  \centering
  \includegraphics[width=0.42\textwidth]{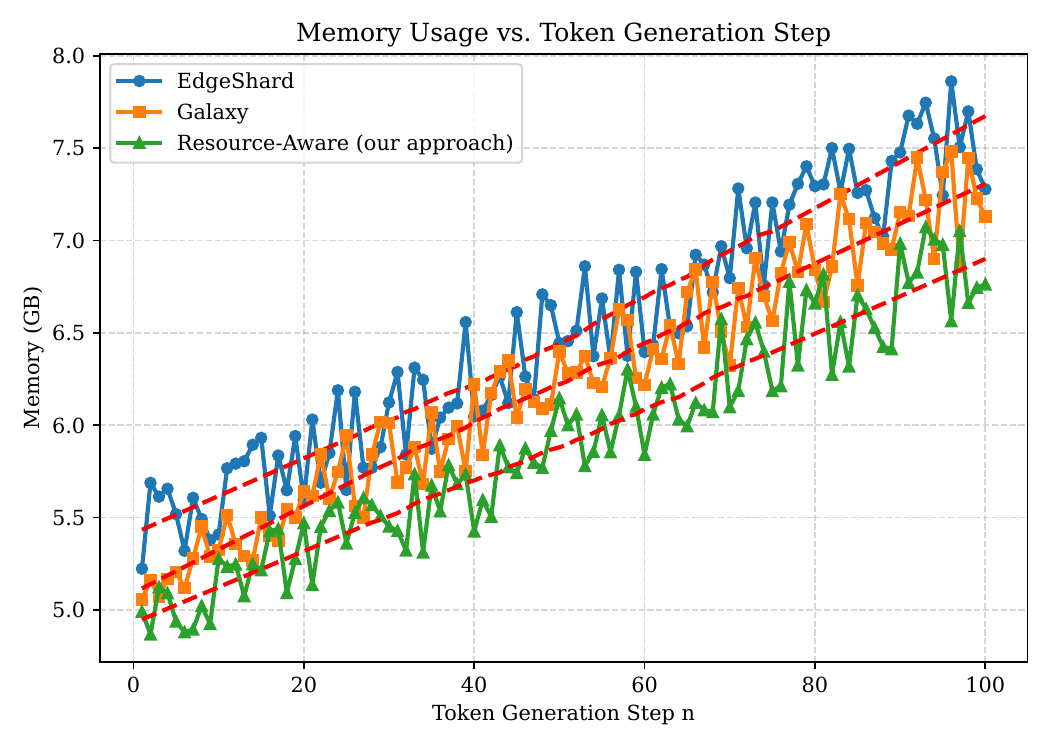}
  \caption{Total memory usage vs.\ generated token step $n$ in the 25-device setup. Our resource-aware approach mitigates memory growth more effectively.}
  \label{fig:mem_big_scenario}
\end{figure}

\paragraph{Memory Usage vs.\ Token Generation Step $n$}
As shown in Figure~\ref{fig:mem_big_scenario}, EdgeShard and Galaxy exceed 7\,GB by $n=100$, while Resource-Aware remains near 6\,GB. The disparity grows further for larger $n$, partly due to the inflexibility of layer-level partitioning under expanding K/V caches.

\paragraph{Scalability with Increasing Number of Devices} Although allocating attention heads among more devices can accelerate inference by spreading the workload, larger networks impose additional coordination and scheduling overhead. As the number of devices grows, the complexity of finding a good placement or migration increases.

\subsection{Summary of Results}
Overall, in the small-scale scenario, our method remains within 15--20\% of the optimal solver’s latency, outperforming simpler heuristics such as the Greedy and Round-Robin baselines by 40--60\%. In the medium-scale setting, our approach scales well to 25 devices and $N=1000$ tokens, resulting in latency and memory overhead considerably better than EdgeShard or Galaxy (up to 9--10 times speedup). This highlights the advantage of \emph{attention head-level} partitioning plus \emph{dynamic} partitioned blocks assignment for autoregressive LLM inference.

\section{Conclusion and Future Work}
\label{section:conclusion}

We introduced a resource-aware approach for partitioning decoder-only Transformers under the tight memory, compute, and link constraints. Our key contribution is to treat each attention head block along with its associated key/value cache as a distinct block that can be allocated to a certain device. The ability to execute attention head blocks in parallel across devices is the central strength of our approach, significantly reducing inference delay. By explicitly modeling how K/V caches expand at each decoding step, we can dynamically reassign these fine-grained blocks across devices to balance workloads. 

For future work, we plan to extend and validate our approach on multi-layer decoder-only Transformers. We also aim to incorporate limited foresight in the decision-making process i.e., to predict resource availability ahead of time and make decisions based on these predictions. We also aim to deploy real-world testbeds that add factors such as energy constraints or inference request load forecasts, in a real environment.

\bibliographystyle{ieeetr}
\bibliography{wiopt2025}

\end{document}